\documentclass[preprint,12pt,authoryear]{elsarticle}
\usepackage{url}
\usepackage{lineno,hyperref}
\usepackage{amsmath}
\usepackage{amssymb}
\usepackage{braket}
\usepackage[super,compress]{cite}
\usepackage{epsfig}
\usepackage{epstopdf}
\usepackage{multirow}
\usepackage{booktabs}
\usepackage{adjustbox}
\usepackage{caption}
\usepackage{array}
\usepackage{verbatim}
\usepackage{stmaryrd}

\usepackage{boondox-cal}
\usepackage{rotating}
\usepackage{natbib}
\usepackage{etoolbox}
\usepackage{boondox-cal}
\usepackage{rotating}
\modulolinenumbers[5]
\journal{New Astronomy}
\usepackage{numcompress}

\begin{document}

\begin{frontmatter}

\title{Beta decay and electron capture rates on manganese isotopes in astrophysical environments}
\author[mymainaddress]{Ramoona Shehzadi\corref{mycorrespondingauthor}}
\cortext[mycorrespondingauthor]{Corresponding author}
\ead{ramoona.physics@pu.edu.pk}

\author[mysecondaryaddress1,mysecondaryaddress2]{Jameel-Un Nabi}
\ead{jameel@uow.edu.pk}

\author[mymainaddress]{Fakeha Farooq}
\ead{fakehafarooq@gmail.com}

\address[mymainaddress]{Department of Physics, University of the Punjab, Lahore 54590, Pakistan}
\address[mysecondaryaddress1]{University of Wah, Quaid Avenue, Wah Cantt 47040, Punjab, Pakistan.}
\address[mysecondaryaddress2]{Faculty of Engineering Sciences, GIK
Institute of Engineering Sciences and Technology, Topi, Khyber Pakhtunkhwa 23640, Pakistan}

\begin{abstract}
The isotopes of manganese in the mass range $A = 53-63$ are abundant
in the core material of high-mass stars and are believed to be
of prime importance in the progression of the pre-collapse phases.
During these late evolutionary phases, nuclear processes associated
with weak interactions, including $\beta^{-}$-decay and electron capture (EC)
on these isotopes, significantly alter the Y$_{e}$
(lepton to baryon ratio) of the core's composition. The temporal
change of this parameter is one of the basic elements to simulate a
successful explosion. Hence, the $\beta^{-}$-decay and EC rates
of manganese (Mn) nuclides may serve as an important input in the simulation codes
of core-collapse supernova. In this paper, we focus on the study
of the weak-decay characteristics of $^{55-63}$Mn nuclides. The strength
distributions of Gamow-Teller transitions in the directions of
$\beta^{-}$-decay and EC for these isotopes were calculated
employing the proton--neutron quasi-particle
random-phase-approximation (pn-QRPA) model. The $\beta$-decay half-life values of Mn
isotopes under terrestrial conditions were computed and compared
with measured data and previous calculations. The pn-QRPA estimated half-lives are in good agreement with the experimental values. The $\beta^{-}$ and EC rates were later calculated covering a large
range of stellar temperatures $(0.01-30)\times 10^{9}\;$K and
densities ($10-10^{11}$) g/cm$^{3}$. The computed rates were compared with previous calculations performed using the
Large-Scale Shell Model (LSSM) and Independent Particle Model (IPM).
For most of the isotopes, in high temperature and
density regimes, the pn-QRPA calculated $\beta^{-}$-decay rates are smaller than the LSSM and IPM rates by up to
an order of magnitude.  The EC rates, on the other hand, are bigger by up to an order of magnitude (factor of $3-10$), at high temperature, when compared with the LSSM (IPM) results.
\end{abstract}

\begin{keyword}
Weak decay rates; core-collapse; Gamow-Teller transitions; pn-QRPA theory
\end{keyword}

\end{frontmatter}

%\linenumbers

%\tableofcontents
%
\section{Introduction}
\label{sec:intro} The most energetic bursts of stars in the universe
are known as supernovae. Among them the gravitational
(core-collapse) supernovae have vital importance in the life cycle
of high-mass stars. They have major contributions to elemental
abundances and thus, to the nucleosynthesis in the
universe~\citep{Woo95,Hash95,Woo02,Heger03,Jan07}. A better
understanding of the driving mechanisms associated with pre- and
post-phases of collapsing cores is required to study and to simulate
their dynamics. It has been long recognised that the weak
interactions, including $\beta^{\pm}$-decays and electron capture
(EC) reactions, have substantial role in the description of
mechanisms involved in the stellar-evolutionary stages prior to the
core-collapse~\citep{Pruet03,Mol03,Rau02,Burbidge57,Wall97}.
Therefore, these processes and associated rates are considered
important inputs for the modelling and simulation studies of
supernovae~\citep{Bethe79,Lang03,Suz16}.

Initially, in hydro-statically equilibrated iron-core, degenerate electron gas supplies the pressure support against the inward self gravity of the core. However, under high stellar temperature
(T $\sim$ 10$^9$ K) and density (10$^{10}$ g/cm$^3$) conditions, the sufficiently high Fermi energy of
degenerate gas favours electron capture on large nuclei and free protons~\citep{Bethe90, Iwa99, Hix03}. EC is responsible for the
neutronisation and deleptonisation of the pre-supernova star. This reduces the electron abundance (Y$_e$) with respect to that of baryons in the core~\citep{Arn77, Bethe79}. The degeneracy pressure of the electron plasma is
not able to counter the core collapse beyond the Chandrasekhar mass limit ($\sim 1.5$ M$_\odot$) and the core starts collapsing with $\sim$1/4 times the velocity of light~\citep{Jos11}. With  further increase in neutronisation, the $\beta^{\pm}$-decay reaction energies (Q$_\beta$) increase and become more favourable
in comparison to EC reactions~\citep{Lang00}. Both $\beta^{\pm}$-decay and EC reactions produce
neutrinos (anti-neutrinos) at pre-supernova densities (up to $\sim$ 10$^{11}$ g/cm$^3$) which leave the stars.
The emission of neutrinos from the stellar interior removes entropy and thus cools the
core~\citep{FulMe91, He01}. Resultantly, both processes strongly impact the Y$_e$, mass and entropy of the core of pre-supernova stars.

The weak processes are largely determined by Gamow-Teller (GT) transitions because
of the existence of extremely high densities and temperatures in
stars~\citep{Bethe79,Fuller80, Fuller82a, Fuller82b,
Fuller85}. The pioneering calculations of weak
rates were done by~\citet{Fuller80, Fuller82a, Fuller82b,
Fuller85} and~\citet{Aufder94} by employing the Independent-Particle Model
(IPM). Later two microscopical approaches, namely; proton-neutron
quasi-particle random-phase-approximation (pn-QRPA)~\citep{Nabi99a,Nabi04} and the Large-Scale Shell Model
(LSSM)~\citep{Lang00,LangMar03}, were introduced for the weak-rates calculations. Although,
LSSM deals with astrophysical rates calculations on a microscopic
level, the formalism is premised on the Brink's
hypothesis~\citep{BAH}. In contrast, in the pn-QRPA model, the GT
strength functions are computed by considering the state-by-state
contribution of each excited state of the decaying nucleus. Another feature of the pn-QRPA model is its large model space (up to 7$\hbar \omega$) and hence it can be applied for the calculation of weak rates for any arbitrary heavy nucleus.

The main constituents of the stellar core during the pre-supernova
evolution are largely pf-shell nuclei~\citep{Aufder94,He01}. These
nuclei control Y$_{e}$ in the final phases of progression after the
silicon burning stages of the core. Therefore, GT characteristics of the
nuclei in this mass region continue to attract attention for
a better comprehension of the stellar weak processes. In the simulation
studies performed by~\citet{Aufder94} and~\citet{Nabi21}, the authors identified the most relevant EC and
$\beta$-decay nuclei that may affect the conditions in pre-supernova
evolution. Among them, isotopes of manganese with
$53 \leq$A$\leq 63$ have significant importance. Regarding
their abundance and weak rates, two sets of Mn isotopes $^{53-61}$Mn
and $^{54-61,63}$Mn were short listed as key EC and $\beta$-decay
nuclei, respectively. Several earlier studies e.g.,~\citet{Majid18},~\citet{Sarrig16},~\citep{Sarrig13} and~\citet{Cole12},
have calculated the $\beta$-decay properties
including GT-strengths, half-lives and weak rates of $^{55}$Mn. In a study performed by~\citet{Poss18},
the authors reported the half-lives of $^{56-69}$Mn.
Recently,~\citet{Azev20}  evaluated the $\beta$-decay rates
and half-lives for isotopes of Mn with 46$\leq$A$\leq$70.

In this work, we
compute the terrestrial $\beta$-decay half-life
values for selected isotopes of manganese and compare them with the experimental half-lives from the Letter of Nuclides~\citeyearpar{LoN21} and previous
theoretical~\citep{Mol19,Poss18,Azev20} data. We estimate the weak rates and B(GT) strengths, in
$\beta^{-}$-decay and EC directions, for $^{55-63}$Mn. We further compare our
rate calculations with the LSSM and IPM rates.  In
Section~\ref{sec:formalism}, we describe a short review of the pn-QRPA
formalism. The results and discussions of our calculations are given
in Section~\ref{sec:results}. The outcome of our findings is
presented in Section~\ref{sec:conclusions}.
\section{Theoretical Formalism}
\label{sec:formalism}
In QRPA theory, states of a system of quasi-particles $(\textit{qp})$ are taken as deformed single-particle $(\textit{sp})$
Nilsson basis. The pairing correlations and residual GT-interactions between paired
nucleons are treated in the BCS approximation and RPA with separable forces,
respectively. The Hamiltonian of the $(\textit{qp})$ system has the form
\begin{equation}
H_{pnQRPA} = H_{sp} + V_{\text{pairing}} + V_{GT_{pp}} + V_{GT_{ph}}~.
\label{Ham}
\end{equation}
The Hamiltonian of the $(\textit{sp})$ system is represented as $H_{sp}$. The state-vectors and energies
of this system are estimated by employing the deformed Nilsson model~\citep{Nil55}.
The pairing correlations and residual GT-interactions are included in the pnQRPA
Hamiltonian as $V_{\text{pairing}}$ and $V_{GT_{pp}}$ ($V_{GT_{ph}}$), respectively.
The nucleons are correlated through constant $V_{\text{pairing}}$ having pairing gaps
$\Delta _{pp}$ ($\Delta _{nn}$) between the protons (neutrons). The proton and neutron separation energies ($S_{p}$ and $S_{n}$) were used to evaluate the pairing gaps as follows~\citep{Wang12}:
\begin{eqnarray}
  \Delta_{pp} &=& \frac{1}{4} (-1)^{Z+1}[S_{p}(A+1, Z+1) - 2S_{p}(A,Z)+S_{p}(A-1,Z-1)], \nonumber \\
  \Delta_{nn} &=& \frac{1}{4} (-1)^{A-Z+1}[S_{n}(A+1, Z) - 2S_{n}(A,Z)+S_{n}(A-1,Z)].
\end{eqnarray}

The GT-residual forces: $V_{GT_{pp}}$ (particle-to-particle) and
$V_{GT_{ph}}$ (particle-to-hole) have force constants $\kappa$ and
$\chi$, respectively. In the present work, experimental
$\beta$-decay half-life values taken from~\citeyearpar{LoN21} were reproduced
through fine tuning of the $\kappa$ and $\chi$ model parameters adopting the 1/A$^{0.7}$ dependence as in Ref.~\citep{Hom96}.
The parameterised expressions of these
constants are
\begin{equation*}
\chi = 1.18/A^{0.7} \;(\text{MeV});~~~~ \kappa = 1.53/A^{0.7} \;(\text{MeV})~.
\end{equation*}

The nuclear deformation parameter ($\beta_{2}$) values used in this work were taken from~\citep{Mol16}.
Other primary parameters of the pn-QRPA model are the
Nilsson oscillator constant ($\Omega = 41/A^{1/3}$\;MeV) and the
Nilsson potential parameters, which were adopted from~\citep{Nil55}. For the present
computations, Q-values were calculated by using experimental values
of mass excess from recent atomic mass evaluation of~\citet{Wang21}.

For a parent-nucleus $i^{th}$ state transition to a daughter-nucleus $j^{th}$ state, the partial decay
rate under stellar conditions is computed by
\begin{eqnarray}
\lambda^{TR}_{ij}=\frac{\ln 2}{D}f_{ij}^{TR}(\rho, E_{f}, T)B_{ij}; ~~~~TR\equiv\beta, EC
\label{Eq:rate}
\end{eqnarray}
The value of constant $D$ is 6143 as reported in~\citet{Har09}. The reduced transition
probability ($B_{ij}$) is the combination of transition probabilities for
GT-decay ($B_{{GT}_{ij}}$) and Fermi decay ($B_{{F}_{ij}}$),
\begin{eqnarray}
\begin{split}
&B_{ij} = \left(\frac{g_{A}}{g_{V}}\right)^{2}B_{GT_{ij}} + B_{F_{ij}}; ~~~~\frac{g_{A}}{g_{V}}=-1.2694,\\
&B_{GT_{ij}} = \frac{|\langle j||\sum_{k}t^{k}_{\pm}\vec{\sigma}^{k}||i \rangle|^{2}}{2I_{i}+1},\\
&B_{F_{ij}} = \frac{|\langle j||\sum_{k}t^{k}_{\pm}||i \rangle|^{2}}{2I_{i}+1}.
\end{split}
\label{Eq:Rprob}
\end{eqnarray}
The value of ratio $g_{A}/g_{V}$ was taken from~\citep{Nak10}. In
Eq.~(\ref{Eq:Rprob}), $I_{i}$, $t^{k}$ and $\vec{\sigma}^{k}$ are
the total $i^{th}$ state spin of nucleus, isospin and spin
operators, respectively. The Fermi integrals $f_{ij}^{TR}(\rho,
E_{f}, T)$ (in natural units) are computed as
\begin{eqnarray}
f^{\beta^{-}}_{ij} = \int_{1}^{\epsilon_{m}}\epsilon({\epsilon^{2} -1)^{1/2}(\epsilon_{m}-\epsilon)^{2}F(+Z,\epsilon)(1-G_{-})}d\epsilon,
\label{Eq:FIEE}
\end{eqnarray}
\begin{eqnarray}
f^{EC}_{ij} = \int _{\epsilon_{l}}^{\infty}\epsilon
(\epsilon^{2} -1)^{1/2}(\epsilon_{m}+\epsilon)^{2}F(+Z,\epsilon)G_{-}\ d\epsilon,
\label{Eq:FIPC}
\end{eqnarray}
where the total energy of the electron ($\epsilon$) is the sum of the kinetic energy
and rest mass energy, $\epsilon_{l}$ ($\epsilon_{m}$) is threshold value of total
EC decay energy (total $\beta$-transition energy). The distribution function
of the electron ($G_{-}$) obeys Fermi-Dirac statistics. The method reported in~\citet{Gove71}
was used for the calculations of the Fermi functions
$F\left(+Z,\epsilon\right)$.

The total decay rate for a nucleus is calculated by summing up the partial rates of all initial
and final levels until the convergence of total rate is achieved.
\begin{equation}
\lambda^{TR}_{total} = \sum _{ij}P_{i} \lambda _{ij}^{TR}; ~~~~TR\equiv\beta, EC
\label{Trate}
\end{equation}
In the above equation it was assumed that the probability of occupation of excited levels of the parent nucleus ($P_{i}$)
obeys the normal Boltzmann distribution.

\section{Results and Discussions} \label{sec:results}

We have calculated GT strength
distribution functions, the terrestrial $\beta$-decay half-life
values and stellar-weak rates (EC and $\beta^{-}$-decay rates) for manganese isotopes with mass
ranging from A $= 55-63$, using the pn-QRPA theory.
As per previous studies, the GT strengths calculated using different nuclear models are generally higher than that measured experimentally (e.g.,~\citet{Gaard83,
Vet89, Roe93, Alford93, Kateb94, William95}). The calculated GT strengths are then renormalised by different models by applying some fixed value of quenching factor. \citet{Vet89} and \citet{Roe93} predicted a quenching factor of 0.6 for the RPA calculation in the case of $^{54}$Fe when comparing their measured strengths to RPA calculations. In the literature we find several studies where authors have used the same quenching factor of 0.6 (including shell model) (e.g.,~\citet{Brown85, Caurier95, Muto91, Oda94, Peterman07, Radha97}).

On the other hand, there are  studies where authors contest that there is no need for introducing an explicit quenching factor (e.g.,~\citet{Ha15, Ozen06}). \citet{Mol97} (FRDM + RPA model) used no quenching factor, \citet{Pfei02} (KHF, QRPA) and \citet{Nabi19, Nabi21} (pn-QRPA) did not employ any quenching factor in their calculations. A recent study of ab-initio calculation~\citet{Gysbers19} suggest that $\beta$-decay rate calculations can be calculated without invoking a quenching factor. This curious finding is likely to attract attention of many theorists performing $\beta$-decay calculations.  In our present calculation, employing a quenching factor of 0.6, to renormalise the BGT strengths, led to a decent comparison of our calculated half-lives with the experimental data. Hence we used the same quenching factor for the computation of weak rates in stellar environment.

In the first part of this section, $\beta$-decay
half-life values of $^{56-63}$Mn isotopes calculated under
terrestrial conditions are presented. Our estimated half-life values
are compared with recent experimental half-lives taken
from~\citeyearpar{LoN21} and a previous calculation of~\citet{Mol19} using QRPA, by ~\citet{Poss18}
and~\citet{Azev20} using different versions of gross theory of beta
decay (namely GTBD1, GTBD2). This comparison is shown in
Table~\ref{table:table1}. It can be seen from Table~\ref{table:table1} that the half-life values of Mn isotopes from present studies
agree rather well with the experimental results. After validating the reliability of our model we proceed to compute the weak rates in stellar environment.

\begin{figure}
	\begin{center}
		\includegraphics[width=0.8\textwidth]{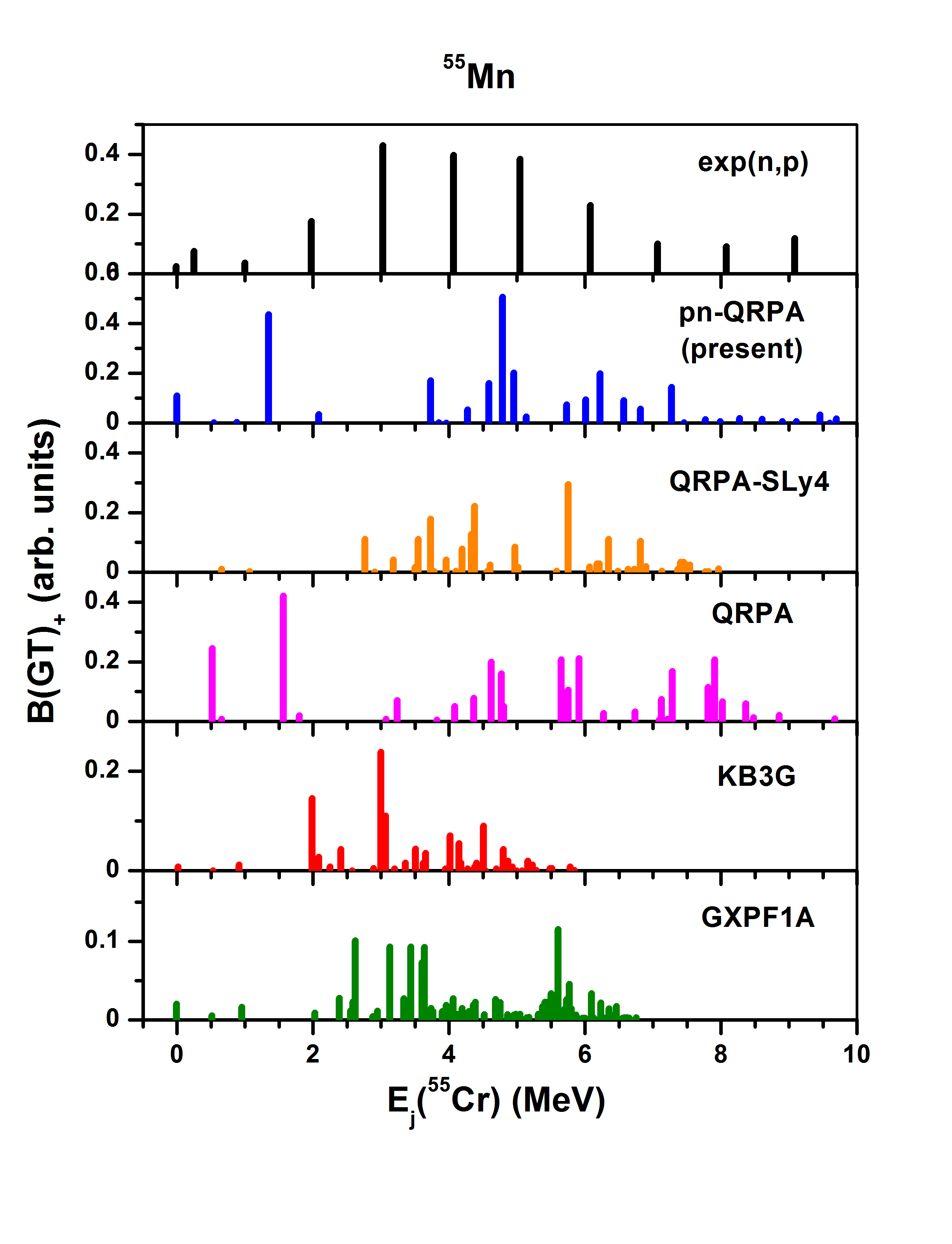}
	\end{center}
	\caption{Comparison of present pn-QRPA calculated B(GT)$_{+}$ strength
		distribution in $^{55}$Mn with measured and previous calculations. GXPF1A~\citep{Honma05} and KB3G~\citep{Poves01}
		show shell model calculations in the full pf-shell space, QRPA represents
		calculation based on the formalism of~\citep{Mol90},
		QRPA-SLy4 shows QRPA calculation with Skyrme force (taken
		from~\citet{Sarrig16}) while exp(n,p) is the experimental data
		from~\citep{Kateb94}.} \label{figure3}
\end{figure}

The allowed GT strength distributions from the ground-level of decaying
nuclei to ground- and excited-levels of residual nuclei for $^{56-59}$Mn and $^{60-63}$Mn
 in the EC direction [B(GT)${_+}$] and in the $\beta^{-}$-decay direction [B(GT)${_-}$] are shown in
Tables~\ref{table:table2} and \ref{table:table3}, respectively. Around 150
daughter excited levels up to an excitation energy of $45\;$MeV are
considered in this study. In Tables.~\ref{table:table2} and~\ref{table:table3},
B(GT)$_\pm$ strengths are given up to $10\;$MeV excitation
energy of the daughter nuclei. In addition to the parent
ground-state, the GT strengths from the parent excited states are determined but not presented here due to space consideration. It is
noted that the excited-state GT strengths differ considerably from
the ground-state GT strengths and discourage the use of so-called Brink's
hypothesis (as used by IPM and LSSM) for the estimation of stellar weak rates. For both ground and excited-states GT strengths, the
ASCII files are available and may be requested from the corresponding
author. For the special case of $^{55}$Mn, the pn-QRPA GT strength distributions
computed in the EC direction [B(GT)$_{+}$] are compared with the data
from (n,p) experiment~\citep{Kateb94} and other theoretical
calculations (see Fig.~\ref{figure3}). The theoretical results of
QRPA and shell-model with GXPF1A and KB3G interactions are taken from~\citet{Cole12}, and the QRPA
calculations are adopted from~\citet{Mol90}. In their work, the authors used ground level
deformation parameter and mass values from the finite droplet model.
The results of B(GT) calculations with GXPF1A interaction~\citep{Honma05} and
KB3G interaction~\citep{Poves01} were achieved within the full
pf-shell model space. The results of QRPA with SLy4 Skyrme force
(QRPA-SLy4) were taken from~\citet{Sarrig16}. It is noted that, in comparison to the shell model
B(GT), the GT strengths computed by the pn-QRPA model are well fragmented.

\begin{figure}
	\begin{center}
		\includegraphics[width=0.9\textwidth]{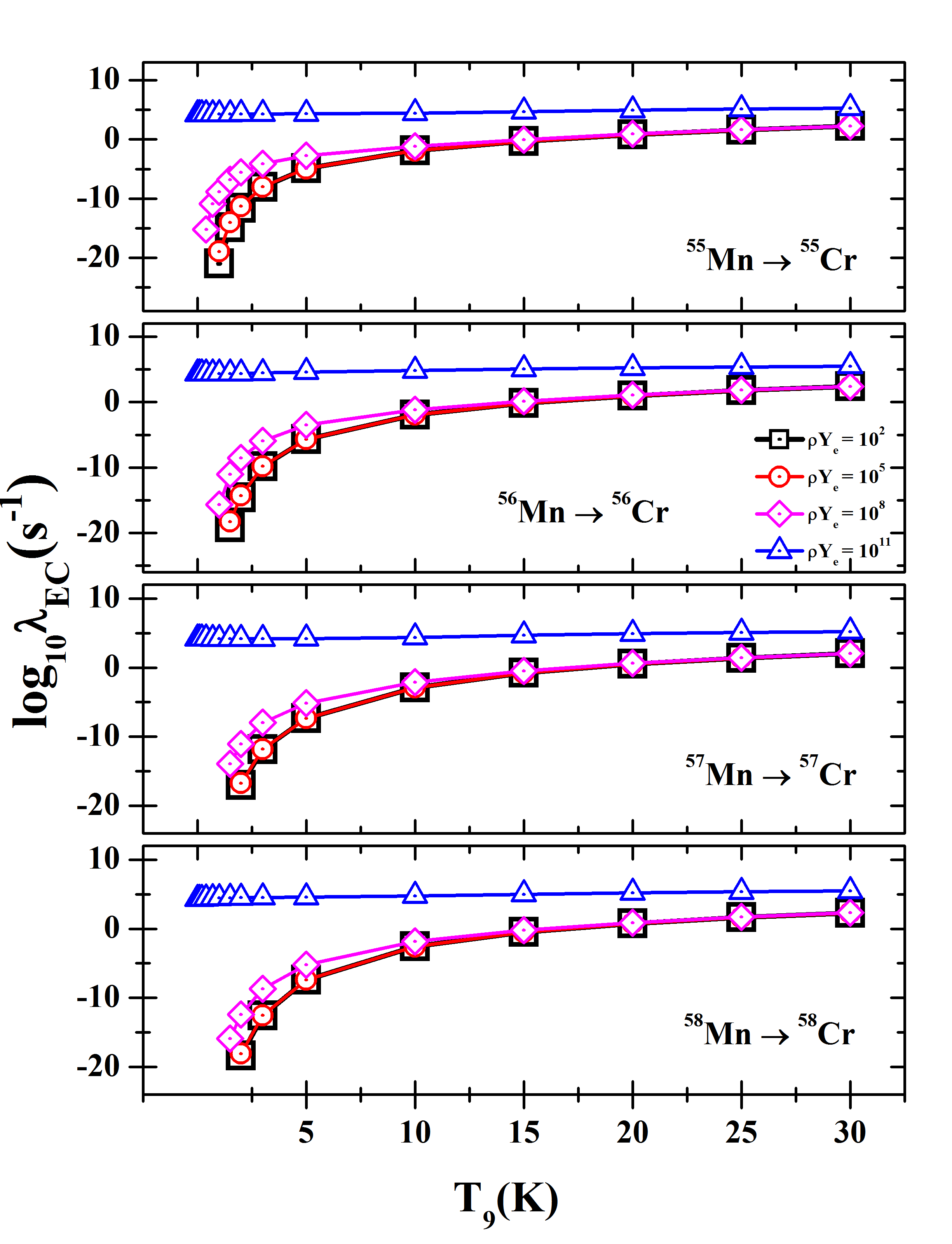}
	\end{center}
	\caption{The pn-QRPA calculated EC rates ($\lambda_{EC}$) for
		$^{55-58}$Mn,  at four selected values of stellar density ($\rho$ Y$_{e}
		= 10^2, 10^{5}, 10^{8}$ and $10^{11}$ in units of g/cm$^{3}$), as a
		function of temperature (T$_{9}$ in units of $10^{9}$\;K.) The
		stellar rates are shown in $\log_{10}$ values and have units of
		s$^{-1}$.} \label{figure4}
\end{figure}

\begin{figure}
	\begin{center}
		\includegraphics[width=0.9\textwidth]{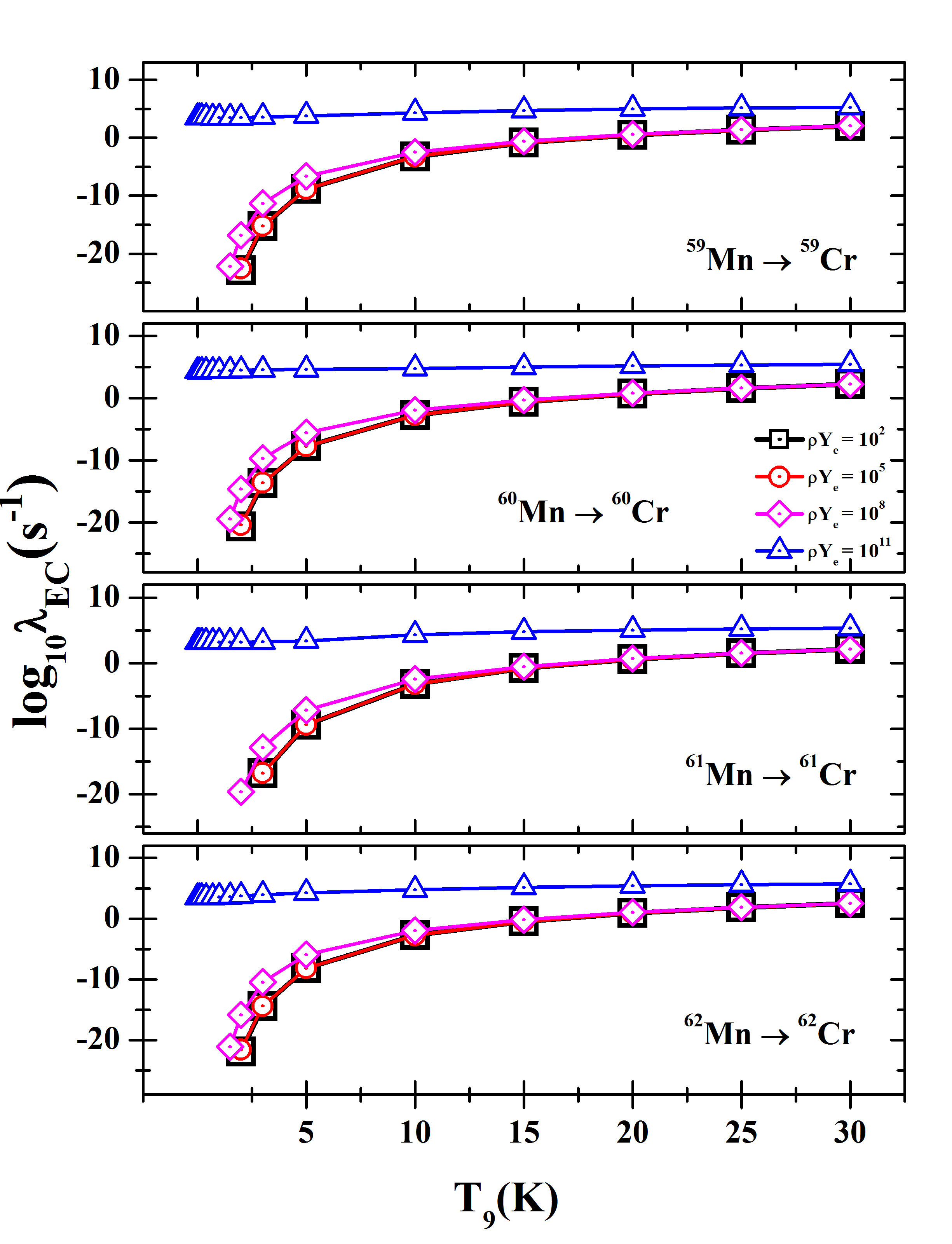}
	\end{center}
	\caption{Same as Fig.~\ref{figure4}, but for $^{59-62}$Mn.}
	\label{figure5}
\end{figure}

\begin{figure}
	\begin{center}
		\includegraphics[width=0.9\textwidth]{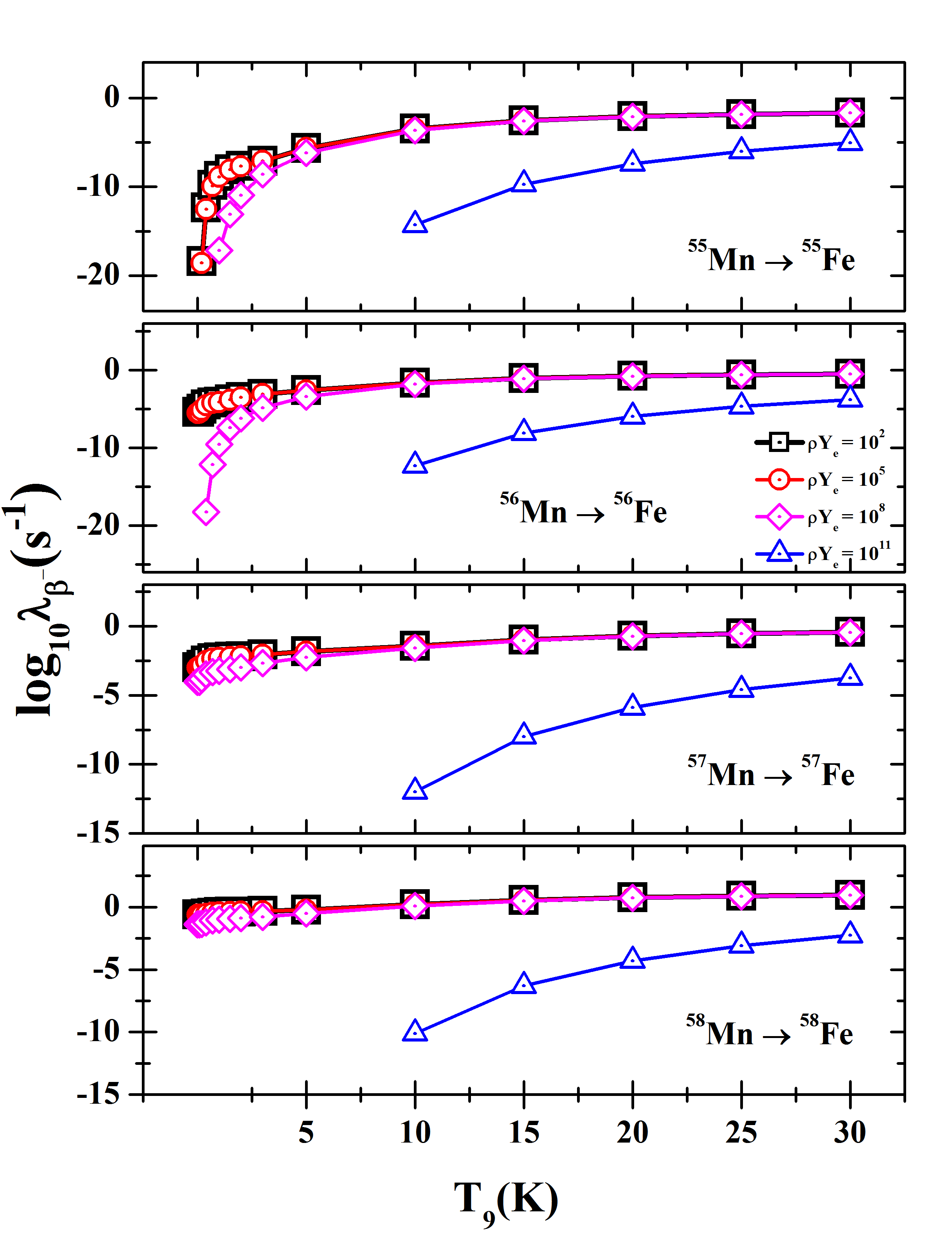}
	\end{center}
	\caption{The pn-QRPA calculated $\beta^{-}$-decay rates
		($\lambda_{\beta^{-}}$) for $^{55-58}$Mn. Other details are the
		same as in Fig.~\ref{figure4}.} \label{figure6}
\end{figure}

\begin{figure}
	\begin{center}
		\includegraphics[width=0.9\textwidth]{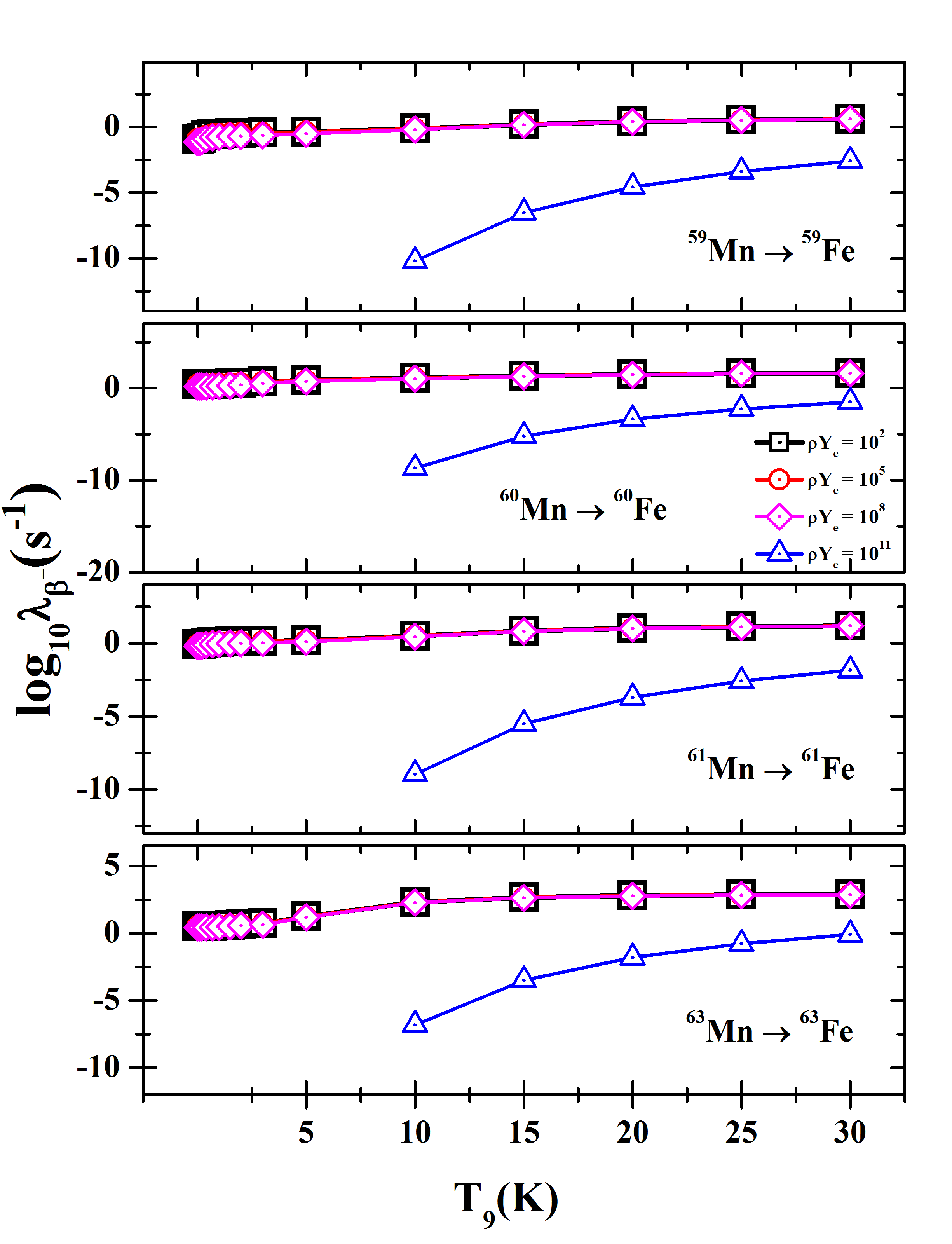}
	\end{center}
	\caption{Same as Fig.~\ref{figure6}, but for $^{59-61,63}$Mn.}\label{figure7}
\end{figure}

Next, we present the pn-QRPA computed EC and $\beta^{-}$-decay weak rates for important manganese isotopes, $^{55-62}$Mn and $^{55-61,63}$Mn, respectively. The weak
rates are computed over a large domain of stellar temperature
$(0.01-30)\times 10^{9}\;$K and density ($10-10^{11}$) g/cm$^{3}$.
The EC rates denoted by $\lambda_{EC}$, for $^{55-62}$Mn isotopes
are shown in Figs.~\ref{figure4} and~\ref{figure5} and the $\beta^{-}$-decay
rates, $\lambda_{\beta^{-}}$, for $^{55-61,63}$Mn are shown in
Figs.~\ref{figure6} and \ref{figure7}. The stellar rates are presented in
$\log_{10}$ values (given in units of s$^{-1}$) as a function of stellar
temperature (T$_{9}$, in units of $10^{9}\;$K) at four values of
stellar density ($\rho$ Y$_{e} = 10^2, 10^{5}, 10^{8}$ and
$10^{11})\;$g/cm$^{3}$. We consider $\rho$ Y$_{e} = 10^2$ as a low density core, $\rho$
Y$_{e} = 10^5, 10^{8}$ as a medium density core and $\rho$ Y$_{e} = 10^{11}$ as a high density core in the following analysis. From Figs.~\ref{figure4}-\ref{figure7}, it can be seen that, for
every Mn isotope, overall EC and $\beta^{-}$-decay rates in each
density region increase with the rise of core temperature. This
increment is primarily due to the expansion of the available phase space with increasing temperature. Additionally, this rise in
temperature causes an increase of the occupation probabilities of
parent excited levels which leads to a larger contribution from
excited states to the total rates. As the stellar core becomes denser, the electron Fermi energy increases which
causes a substantial increase (decrease) of EC ($\beta^{-}$-decay)
rates at high values of stellar density (as can be seen from
Figs.~\ref{figure4}-\ref{figure7}).

Finally we present the comparison of our weak rates with the IPM~\citet{Fuller80, Fuller82a, Fuller82b,
Fuller85} and LSSM~\citet{Lang01} calculations.
For the purpose of comparison, ratios are
computed between current and previous rates (wherever available).
Table~\ref{table:table4} presents the ratios of pn-QRPA computed EC rates to
the corresponding LSSM and IPM rates at three selected values of density ($\rho$ Y$_{e} = 10^3,
10^{7}, 10^{11}$)g/cm$^{3}$ as a function of core temperature. Table~\ref{table:table4}
shows that for most of the temperature and density values, our estimated weak
rates of EC interaction are generally bigger than the rates calculated by LSSM by a factor 2
and up to one order of magnitude. For comparison with IPM rates, it
may be noted from Table~\ref{table:table4} that, at $\rho$ Y$_{e} = 10^3
~\text{and}~ 10^{7}$,  the pn-QRPA calculated rates are bigger by
up to  2 orders of magnitude. It is further noted that at high density ($\rho$ Y$_{e} = 10^{11}$), for temperatures T$_{9}
< 30$, for some of the isotopes (e.g., $^{58, 60}$Mn) the rates from the two
model calculations are comparable to each other. For the remaining isotopes,
within this temperature range, the IPM rates are bigger than our
calculated rates by a factor of 2--6. However, at T$_{9} = 30$, the
pn-QRPA rates get enhanced by a factor of 3--10. The
reason for this difference between the calculated rates using
different models may be attributed to the use of Brink's hypothesis
in case of LSSM and IPM calculations in order to estimate excited state GT
transition strengths. In contrast, the pn-QRPA model performs the
state-by-state estimation of GT transition strengths for the
computation of stellar rates. The comparison of $\beta^{-}$-decay
rates with the LSSM and IPM calculations is shown in
Table~\ref{table:table5}. For $^{55,56}$Mn, our computed rates are,
in general, smaller than the rates determined by LSSM (IPM) by up to 1 (2) order of magnitude,
for most of the temperature and density ranges. In case
of $^{57}$Mn, at $\rho$ Y$_{e} = 10^3 ~\text{and}~ 10^{7}$, the values of
$\lambda_{\beta^{-}}$ from pn-QRPA and LSSM calculations are comparable
within a factor of $2-3$ and at $\rho$ Y$_{e} = 10^{11}$, the LSSM rates surpass the
pn-QRPA rate values by up to an order of magnitude. The IPM and pn-QRPA rates for this isotope
are comparable at lower temperatures (T$_{9} < 5$), however IPM rates are larger with respect to
the pn-QRPA computed rates by up to an order of magnitude at higher temperatures.
For $^{58,59,60}$Mn, in general,  for most of the temperature and density values, the pn-QRPA results are
enhanced than those of LSSM by a factor 2 up to an order of magnitude. For these
three isotopes, at lower temperatures (T$_{9} < 10$) and densities ($\rho$
Y$_{e} = 10^3 ~\text{and}~ 10^{7}$), the rates computed by pn-QRPA model are larger by factor
$\sim 2 - 9$ than the IPM rates. At higher temperature and
density the IPM rates get enhanced by a factor 2 and up to
an order of magnitude. The convergence issues noted in LSSM~\citep{Pruet03}) and the misplacement of GT centroid (see~\citep{Lang00}), approximations applied
to the unmeasured nuclear matrix elements and no quenching applied to GT
strengths in IPM calculation are other sources of differences between current and previous calculations.

\section{Conclusions}
\label{sec:conclusions}

For a better comprehension of the dynamics of stellar progression and the
core-collapse of high-mass stars, reliable and microscopic calculations of
weak rates are essential. The pn-QRPA theory has a good
track record for the calculation of $\beta$-decay half-lives~\citep{Hir93, Sta90}. In contrast to previous calculations (e.g.,LSSM and IPM), the pn-QRPA theory does not utilise Brink's hypothesis and has access to a large model space.
The pn-QRPA model employed in the current study took the nuclear deformation into
consideration. The GT transition strengths and $\beta$-decay half-lives for selected manganese isotopes ($55 \le A \le 63$) were computed. From simulation
studies, the Mn isotopes in this mass range are considered to be important
from an astrophysical point of view. The EC and $\beta$-decay rates were
calculated over a wide range of stellar temperature and density scales. The computed half-lives were in good agreement with the measured data.
The weak rates of selected Mn isotopes, under stellar conditions,  were later computed and compared
with previous calculations. The reported stellar $\beta^{-}$-decay
rates are smaller than the corresponding IPM and
LSSM rates. On the other hand, the pn-QRPA calculated stellar EC rates
are bigger than the previously assumed rates. Our calculated rates may be used as a nuclear physics input for the core-collapse simulation codes and may result in some interesting outcome. 
%We will urge core-collapse simulators to test run our reported weak
%interaction rates presented here to check their effect on the simulation results. }

\begin{table}[pt]
\caption{\small Comparison of the pn-QRPA calculated terrestrial
half-lives of manganese isotopes with experimental
data~\citep{LoN21} and previous calculations using Gross theory~\citep{Poss18}
$\&$~\citep{Azev20} and QRPA~\citep{Mol19}. }\label{table:table1} \hspace{-1cm} {\small
\centering {\resizebox{!}{3cm}{%
\begin{tabular}{ccccccc}%{\textwidth}{c @{\extracolsep{\fill}} ccccccc}
& & & & & &  \\
\toprule
\multirow{2}{*}{Nuclei} & \multicolumn{6}{c} {Half-life (seconds)} \\
\cmidrule{2-3} \cmidrule{3-4}  \cmidrule{5-6}  \cmidrule{6-7} &
\multicolumn{1}{c}{Exp.} & \multicolumn{1}{c}{pn-QRPA} &
\multicolumn{1}{c}{QRPA (M\"{o}ller)} & \multicolumn{1}{c}{GTBD1\citeyear{Poss18}} &
\multicolumn{1}{c}{GTBD1~\citeyear{Azev20}}  & \multicolumn{1}{c}{GTBD2~\citeyear{Azev20}}\\
\midrule
$^{56}$Mn  &   9284.04$\pm$0.36  & 10636.32  &  --  &  9.74 &  10  &  18022  \\
$^{57}$Mn  &   85.40$\pm$1.80   &  86.86  &  46.35  &  54.28 &  55.49  &  248.51   \\
$^{58}$Mn  &   3.00$\pm$0.10   &  3.12  &  1.24  &  0.92  &  0.95  &  10.44  \\
$^{59}$Mn  &   4.59$\pm$0.05   &  5.66 &  1.06  &  2.54  &  22.57  &  74.88  \\
$^{60}$Mn  &   0.28$\pm$0.02   &  0.29  &  0.44 &  0.35  &  0.42  &  4.07   \\
$^{61}$Mn  &   0.71$\pm$0.008   &  0.85  &  0.21  &  0.74  &  0.77  &  1.28   \\
$^{62}$Mn  &   0.09$\pm$0.013   &  0.07  &  0.06  &  0.18  &  0.20  &  0.22  \\
$^{63}$Mn  &   0.28$\pm$0.004   &  0.20  &  0.06  &  0.39  &  0.40  &  0.27 \\
 \bottomrule
\end{tabular}}}}
\end{table}

\begin{table}[pt]
\caption{\small pn-QRPA calculated GT strength distributions for EC [B(GT)+] and $\beta^{-}$-decay [B(GT)-], for
Mn isotopes, as a function of daughter excitation energies [E$_j$].}\label{table:table2} \hspace{-3cm} {\small
\centering {\scalebox{0.75}{
\begin{tabular}{|c|c|c|c|c|c|c|c|c|c|c|c|}
\hline
\multicolumn{3}{|c|}{Mn56} & \multicolumn{3}{c|}{Mn57} & \multicolumn{3}{c|}{Mn58} & \multicolumn{3}{c|}{Mn59}\tabularnewline
\hline
\hline
Ej & B(GT)+ & B(GT)- & Ej & B(GT)+ & B(GT)- & Ej & B(GT)+ & B(GT)- & Ej & B(GT)+ & B(GT)-\tabularnewline
\hline
(MeV) & (arb. units) & (arb. units) & (MeV) & (arb. units) & (arb. units) & (MeV) & (arb. units) & (arb. units) & (MeV) & (arb. units) & (arb. units)\tabularnewline
\hline
0.85 & 1.12E-01 & 1.59E-03 & 0.01 & 8.39E-02 & 9.44E-03 & 0.00 & 8.07E-04 & 1.81E-02 & 0.00 & 8.33E-03 & 4.25E-02\tabularnewline
\hline
2.09 & 2.14E-01 & 1.06E-01 & 0.14 & 4.72E-01 & 5.37E-03 & 0.81 & 3.74E-01 & 3.10E-02 & 0.47 & 3.25E-01 & 3.42E-02\tabularnewline
\hline
2.66 & 4.44E-02 & 6.06E-01 & 0.37 & 4.52E-04 & 8.12E-03 & 1.67 & 5.10E-01 & 6.00E-01 & 1.95 & 1.29E-01 & 1.36E-03\tabularnewline
\hline
2.96 & 3.13E-02 & 3.09E-02 & 0.86 & 2.67E+00 & 0.00E+00 & 2.26 & 3.47E+00 & 1.41E-01 & 2.51 & 4.95E-03 & 1.40E-02\tabularnewline
\hline
4.98 & 1.79E-01 & 4.02E-01 & 1.36 & 1.72E-01 & 3.06E-03 & 2.78 & 2.84E-02 & 2.27E-02 & 3.61 & 1.50E-04 & 1.52E-03\tabularnewline
\hline
5.10 & 1.46E+00 & 2.63E+00 & 1.68 & 3.51E-02 & 6.12E-02 & 3.99 & 4.04E-02 & 1.35E+00 & 3.84 & 2.54E-02 & 3.03E-03\tabularnewline
\hline
5.71 & 2.29E+00 & 0.00E+00 & 3.99 & 1.30E-02 & 1.27E-02 & 4.82 & 1.86E-01 & 2.22E-03 & 4.14 & 2.41E-01 & 8.28E-03\tabularnewline
\hline
6.12 & 2.55E-02 & 4.45E-02 & 4.10 & 3.03E-05 & 1.73E-06 & 5.00 & 2.47E-02 & 2.58E-01 & 4.48 & 1.38E-02 & 2.75E-04\tabularnewline
\hline
6.30 & 1.06E-03 & 4.49E-02 & 4.98 & 7.02E-02 & 1.03E-01 & 5.22 & 2.05E-03 & 1.55E-05 & 4.86 & 2.66E-01 & 2.49E-02\tabularnewline
\hline
6.56 & 5.98E-02 & 3.84E-02 & 5.11 & 3.45E-05 & 1.21E-02 & 5.50 & 6.38E-02 & 7.45E-02 & 5.10 & 9.52E-02 & 6.67E-03\tabularnewline
\hline
6.76 & 4.40E-01 & 7.47E-02 & 5.26 & 1.25E-01 & 2.96E-02 & 5.84 & 2.32E-02 & 1.10E-01 & 5.48 & 5.66E-02 & 1.03E-01\tabularnewline
\hline
6.89 & 1.90E-01 & 3.98E-01 & 5.52 & 1.15E-01 & 6.56E-02 & 6.53 & 2.02E-01 & 1.98E-01 & 5.60 & 3.21E-02 & 2.32E-02\tabularnewline
\hline
7.07 & 4.20E-01 & 2.47E-01 & 5.85 & 8.34E-02 & 4.09E-02 & 6.64 & 6.44E-02 & 1.28E-03 & 5.72 & 1.69E-01 & 1.68E-02\tabularnewline
\hline
7.29 & 6.79E-06 & 1.73E-04 & 6.16 & 3.99E-02 & 1.90E-03 & 6.77 & 9.75E-04 & 2.52E-01 & 5.90 & 5.16E-04 & 4.95E-06\tabularnewline
\hline
7.63 & 5.63E-02 & 4.57E-01 & 6.57 & 1.59E-02 & 1.22E-04 & 6.87 & 1.51E-02 & 3.54E-02 & 6.12 & 5.47E-02 & 7.77E-03\tabularnewline
\hline
8.21 & 9.46E-02 & 1.71E-03 & 6.74 & 8.33E-02 & 5.72E-01 & 7.19 & 4.87E-02 & 2.51E-01 & 6.67 & 1.55E-01 & 1.08E-01\tabularnewline
\hline
8.41 & 1.46E-01 & 6.89E-02 & 6.91 & 6.84E-02 & 3.21E-01 & 7.59 & 2.08E-02 & 8.55E-05 & 6.97 & 8.24E-03 & 1.55E-02\tabularnewline
\hline
8.55 & 1.25E-01 & 6.43E-02 & 7.26 & 2.49E-03 & 5.35E-01 & 7.86 & 1.10E-01 & 1.92E-01 & 7.11 & 5.63E-02 & 5.33E-01\tabularnewline
\hline
8.65 & 2.87E-04 & 4.20E-02 & 7.41 & 6.25E-03 & 3.35E-01 & 8.03 & 1.96E-03 & 5.92E-01 & 7.29 & 2.95E-02 & 2.57E-01\tabularnewline
\hline
8.98 & 8.09E-02 & 7.77E-02 & 7.81 & 1.07E-02 & 1.84E-01 & 8.16 & 2.25E-05 & 6.78E-03 & 7.45 & 3.93E-02 & 2.39E-03\tabularnewline
\hline
9.15 & 6.64E-02 & 2.38E-03 & 8.19 & 2.99E-02 & 4.29E-02 & 8.48 & 1.17E-03 & 7.78E-01 & 7.66 & 1.81E-02 & 2.71E-01\tabularnewline
\hline
9.55 & 2.68E-02 & 5.54E-04 & 8.30 & 4.30E-02 & 2.50E-01 & 8.63 & 3.10E-03 & 3.03E-01 & 7.85 & 1.41E-04 & 4.98E-01\tabularnewline
\hline
9.74 & 1.21E-01 & 2.06E-02 & 8.50 & 1.99E-02 & 1.81E-04 & 9.08 & 1.45E-02 & 4.59E-02 & 8.15 & 6.92E-02 & 2.42E-02\tabularnewline
\hline
10.18 & 1.40E-02 & 5.79E-03 & 8.69 & 9.64E-03 & 4.01E-03 & 9.30 & 1.00E-03 & 8.64E-04 & 8.27 & 7.21E-02 & 4.65E-02\tabularnewline
\hline
-- & -- & -- & 8.80 & 3.16E-02 & 2.42E-02 & 9.52 & 2.66E-02 & 2.41E-01 & 8.41 & 5.79E-03 & 1.89E-03\tabularnewline
\hline
-- & -- & -- & 9.08 & 1.08E-03 & 3.18E-01 & 9.90 & 1.59E-02 & 5.53E-02 & 8.87 & 2.00E-03 & 1.06E-01\tabularnewline
\hline
-- & -- & -- & 9.20 & 1.39E-03 & 2.07E-01 & 10.00 & 6.80E-03 & 2.94E-02 & 9.01 & 5.89E-03 & 1.08E-01\tabularnewline
\hline
-- & -- & -- & 9.44 & 3.91E-03 & 3.16E-02 & -- & -- & -- & 9.13 & 1.68E-05 & 2.12E-02\tabularnewline
\hline
-- & -- & -- & 9.73 & 1.86E-05 & 4.31E-02 & -- & -- & -- & 9.36 & 3.68E-03 & 6.59E-02\tabularnewline
\hline
-- & -- & -- & 9.87 & 1.70E-02 & 8.06E-03 & -- & -- & -- & 9.50 & 1.20E-03 & 6.35E-01\tabularnewline
\hline
-- & -- & -- & 10.12 & 1.12E-03 & 4.84E-03 & -- & -- & -- & 9.65 & 1.46E-05 & 1.71E-03\tabularnewline
\hline
-- & -- & -- & -- & -- & -- & -- & -- & -- & 9.87 & 4.45E-03 & 8.29E-03\tabularnewline
\hline
-- & -- & -- & -- & -- & -- & -- & -- & -- & 10.03 & 3.46E-02 & 4.68E-02\tabularnewline
\hline
\end{tabular}}}}
\end{table}

\begin{table}[pt]
\caption{\small Same as in Table~\ref{table:table2}, but for $^{60-63}$Mn.}\label{table:table3} \hspace{-3cm} {\small
\centering {\scalebox{0.75}{
\begin{tabular}{|c|c|c|c|c|c|c|c|c|c|c|c|}
\hline
\multicolumn{3}{|c|}{Mn60} & \multicolumn{3}{c|}{Mn61} & \multicolumn{3}{c|}{Mn62} & \multicolumn{3}{c|}{Mn63}\tabularnewline
\hline
\hline
Ej & B(GT)+ & B(GT)- & Ej & B(GT)+ & B(GT)- & Ej & B(GT)+ & B(GT)- & Ej & B(GT)+ & B(GT)-\tabularnewline
\hline
(MeV) & (arb. units) & (arb. units) & (MeV) & (arb. units) & (arb. units) & (MeV) & (arb. units) & (arb. units) & (MeV) & (arb. units) & (arb. units)\tabularnewline
\hline
0.00 & 3.82E+00 & 0.00E+00 & 0.00 & 1.71E-01 & 8.22E-02 & 0.00 & 2.57E-01 & 3.84E-02 & 0.00 & 1.19E-01 & 8.92E-02\tabularnewline
\hline
0.82 & 4.00E-01 & 1.39E-02 & 1.06 & 1.25E-01 & 4.03E-03 & 0.88 & 2.55E-01 & 4.59E-01 & 0.70 & 8.22E-03 & 5.96E-02\tabularnewline
\hline
1.97 & 2.68E-01 & 2.60E-01 & 1.24 & 3.43E-04 & 2.61E-02 & 2.39 & 2.58E-01 & 1.43E-01 & 3.10 & 3.09E-02 & 6.56E-02\tabularnewline
\hline
2.36 & 4.98E-01 & 5.16E-03 & 3.55 & 1.88E-01 & 3.19E-03 & 4.49 & 1.11E-03 & 3.40E-02 & 3.38 & 6.42E-03 & 1.72E-02\tabularnewline
\hline
2.71 & 3.62E-03 & 9.34E-01 & 3.68 & 4.66E-02 & 2.68E-03 & 4.63 & 4.05E-01 & 1.11E-02 & 3.69 & 2.69E-02 & 8.23E-02\tabularnewline
\hline
3.04 & 2.21E-01 & 2.86E-01 & 3.92 & 6.21E-02 & 1.23E-02 & 5.12 & 1.97E-03 & 9.04E-03 & 3.85 & 6.65E-04 & 3.84E-02\tabularnewline
\hline
4.73 & 7.62E-02 & 4.06E-02 & 4.26 & 1.70E-02 & 1.26E-03 & 5.57 & 1.40E-01 & 5.31E-02 & 4.01 & 1.05E-03 & 1.94E-02\tabularnewline
\hline
5.10 & 1.11E-02 & 8.57E-03 & 4.39 & 8.60E-02 & 5.71E-03 & 5.72 & 1.40E-03 & 1.86E-02 & 4.63 & 1.96E-01 & 1.78E-01\tabularnewline
\hline
5.33 & 3.07E-06 & 2.78E-03 & 4.51 & 2.37E-02 & 3.27E-03 & 6.01 & 1.34E-01 & 4.31E-02 & 5.08 & 8.42E-02 & 9.61E-02\tabularnewline
\hline
5.54 & 4.91E-02 & 4.33E-04 & 4.88 & 1.50E-03 & 3.73E-07 & 6.32 & 7.51E-02 & 2.83E-02 & 5.53 & 1.01E-03 & 1.08E-02\tabularnewline
\hline
6.03 & 1.10E-01 & 2.06E-01 & 5.36 & 5.57E-03 & 2.85E-02 & 6.49 & 1.51E-01 & 9.13E-02 & 5.86 & 3.89E-04 & 1.07E-02\tabularnewline
\hline
6.36 & 1.48E-02 & 5.37E-04 & 5.51 & 2.23E-02 & 3.79E-04 & 6.88 & 2.56E-04 & 4.57E-03 & 6.09 & 3.73E-03 & 1.95E-03\tabularnewline
\hline
6.47 & 1.90E-02 & 9.75E-02 & 5.79 & 1.96E-01 & 1.97E-02 & 7.14 & 4.99E-03 & 4.06E-04 & 6.27 & 3.03E-03 & 6.40E-03\tabularnewline
\hline
6.60 & 1.90E-01 & 1.89E-01 & 5.93 & 9.26E-02 & 3.77E-02 & 7.37 & 5.10E-05 & 1.37E-02 & 6.41 & 7.31E-03 & 1.59E-01\tabularnewline
\hline
6.87 & 6.86E-04 & 1.36E-01 & 6.30 & 8.51E-02 & 7.35E-02 & 7.50 & 5.61E-02 & 1.42E-01 & 6.52 & 1.41E-02 & 9.78E-03\tabularnewline
\hline
7.07 & 1.47E-02 & 3.39E-02 & 6.41 & 7.97E-03 & 6.26E-02 & 7.91 & 6.89E-05 & 1.30E-02 & 6.70 & 1.71E-04 & 6.88E-03\tabularnewline
\hline
7.24 & 3.35E-03 & 1.85E-02 & 6.57 & 6.85E-02 & 3.35E-02 & 8.13 & 2.62E-01 & 2.40E-01 & 6.97 & 4.01E-04 & 1.43E-03\tabularnewline
\hline
7.41 & 6.35E-04 & 1.90E-03 & 6.72 & 3.55E-03 & 4.72E-03 & 8.33 & 1.65E-05 & 2.73E-05 & 7.11 & 1.49E-03 & 9.48E-05\tabularnewline
\hline
7.81 & 1.82E-02 & 1.37E-03 & 6.86 & 3.84E-02 & 7.55E-03 & 8.43 & 1.34E-03 & 1.77E-02 & 7.42 & 3.92E-03 & 2.38E-01\tabularnewline
\hline
8.16 & 1.91E-04 & 7.94E-04 & 7.05 & 4.11E-04 & 5.29E-04 & 8.54 & 2.85E-03 & 6.04E-01 & 7.60 & 1.26E-02 & 6.01E-03\tabularnewline
\hline
8.33 & 4.29E-03 & 3.14E-02 & 7.43 & 1.97E-01 & 1.46E-01 & 8.77 & 9.92E-02 & 1.26E+00 & 7.84 & 9.80E-03 & 3.75E-02\tabularnewline
\hline
8.52 & 2.71E-02 & 2.15E-01 & 7.54 & 8.44E-04 & 1.98E-02 & 9.24 & 4.29E-03 & 1.05E-01 & 8.18 & 3.19E-03 & 6.35E-01\tabularnewline
\hline
8.72 & 6.41E-03 & 6.29E-01 & 7.78 & 1.18E-03 & 4.23E-03 & 9.44 & 8.25E-03 & 8.95E-02 & 8.31 & 2.37E-04 & 1.56E-01\tabularnewline
\hline
8.87 & 1.25E-02 & 3.02E-01 & 7.98 & 4.85E-02 & 4.37E-01 & 9.58 & 1.39E-02 & 1.73E-01 & 8.49 & 3.68E-05 & 1.48E-02\tabularnewline
\hline
9.06 & 3.34E-03 & 2.51E-03 & 8.10 & 6.32E-03 & 2.17E-01 & 9.76 & 1.71E-02 & 6.97E-02 & 8.64 & 5.82E-02 & 1.22E+00\tabularnewline
\hline
9.26 & 5.03E-03 & 9.94E-01 & 8.23 & 4.90E-03 & 9.82E-02 & 9.88 & 7.21E-03 & 4.52E-01 & 8.86 & 4.65E-02 & 4.05E-01\tabularnewline
\hline
9.37 & 6.06E-04 & 3.19E-01 & 8.46 & 1.14E-03 & 1.66E-01 & 10.16 & 6.75E-03 & 1.55E-01 & 9.03 & 5.10E-03 & 1.56E-01\tabularnewline
\hline
9.72 & 1.01E-02 & 1.34E-02 & 8.63 & 1.65E-02 & 2.27E-01 & -- & -- & -- & 9.23 & 8.83E-03 & 9.48E-01\tabularnewline
\hline
9.90 & 2.07E-02 & 1.89E-01 & 8.78 & 3.61E-03 & 7.51E-01 & -- & -- & -- & 9.48 & 4.87E-03 & 2.24E-02\tabularnewline
\hline
10.15 & 7.86E-04 & 1.15E-01 & 9.16 & 9.70E-02 & 7.48E-02 & -- & -- & -- & 9.64 & 2.61E-02 & 1.60E-02\tabularnewline
\hline
-- & -- & -- & 9.30 & 8.20E-04 & 6.11E-02 & -- & -- & -- & 10.00 & 4.68E-04 & 1.31E-02\tabularnewline
\hline
-- & -- & -- & 9.76 & 3.58E-03 & 1.46E-01 & -- & -- & -- & -- & -- & --\tabularnewline
\hline
 &  &  & 10.02 & 5.87E-03 & 4.39E-01 & -- & -- & -- & -- & -- & --\tabularnewline
\hline
\end{tabular}}}}
\end{table}

\begin{table}[pt]
\caption{\small The ratios of pn-QRPA EC rates to  shell model rates~\citep{Lang00} [R$_{\text{EC}}$(QRPA/LSSM)]
and to independent particle model rates~\citep{Fuller80, Fuller82a, Fuller82b,
Fuller85} [R$_{\text{EC}}$(QRPA/IPM)]  for
$^{55-60}$Mn. The ratios are calculated at selected values
of stellar temperature, T$_{9}$ (in units of $10^{9}\;$K) and
density, $\rho$Y$_{e}$ (in units of
g$\;$cm$^{-3}$).}\label{table:table4} \hspace{-1cm}
\setlength{\aboverulesep}{0pt} \setlength{\belowrulesep}{0pt}
{\small
\centering {%\resizebox{!}{5cm}{%
\begin{tabular}{|cccccccc|}%{\textwidth}{c @{\extracolsep{\fill}} ccccccc}
\toprule
{\rule{0pt}{13pt} T$_{9}$} & $\rho$ Y$_{e}$ & $^{55}$Mn & $^{56}$Mn & $^{57}$Mn & $^{58}$Mn & $^{59}$Mn & $^{60}$Mn \\[0.4ex]
\midrule
\midrule
\multicolumn{8}{|c|}{\rule{0pt}{12pt} R$_{\text{EC}}$(QRPA/LSSM) } \\[0.4ex]
\midrule
3 & $10^{3}$ & 1.31E+01 & 1.71E-02 & 3.73E+00 & 7.38E-02 & 1.06E+01 & 3.85E+00\tabularnewline
5 & $10^{3}$ & 1.46E+01 & 1.25E+00 & 2.81E+00 & 1.36E+00 & 1.09E+01 & 2.09E+01\tabularnewline
10 & $10^{3}$ & 7.35E+00 & 1.04E+01 & 4.52E+00 & 1.25E+01 & 1.60E+01 & 4.68E+01\tabularnewline
30 & $10^{3}$ & 1.15E+01 & 2.32E+01 & 1.46E+01 & 3.89E+01 & 3.24E+01 & 9.18E+01\tabularnewline
 &  &  &  &  &  &  & \tabularnewline
3 & $10^{7}$ & 1.35E+01 & 1.79E-02 & 3.79E+00 & 7.46E-02 & 1.07E+01 & 3.88E+00\tabularnewline
5 & $10^{7}$ & 1.50E+01 & 1.27E+00 & 2.84E+00 & 1.37E+00 & 1.10E+01 & 2.10E+01\tabularnewline
10 & $10^{7}$ & 7.36E+00 & 1.04E+01 & 4.53E+00 & 1.25E+01 & 1.60E+01 & 4.68E+01\tabularnewline
30 & $10^{7}$ & 1.15E+01 & 2.33E+01 & 1.46E+01 & 3.89E+01 & 3.23E+01 & 9.18E+01\tabularnewline
 &  &  &  &  &  &  & \tabularnewline
1 & $10^{11}$ & 1.69E+00 & 2.75E+00 & 2.65E+00 & 7.41E+00 & 1.16E+00 & 8.79E+00\tabularnewline
1.5 & $10^{11}$ & 1.72E+00 & 2.67E+00 & 2.61E+00 & 7.89E+00 & 1.16E+00 & 1.08E+01\tabularnewline
2 & $10^{11}$ & 1.75E+00 & 2.75E+00 & 2.60E+00 & 8.22E+00 & 1.19E+00 & 1.24E+01\tabularnewline
3 & $10^{11}$ & 1.82E+00 & 3.19E+00 & 2.61E+00 & 8.71E+00 & 1.43E+00 & 1.48E+01\tabularnewline
5 & $10^{11}$ & 1.95E+00 & 4.38E+00 & 2.72E+00 & 9.73E+00 & 2.31E+00 & 1.79E+01\tabularnewline
10 & $10^{11}$ & 2.39E+00 & 7.23E+00 & 3.89E+00 & 1.46E+01 & 7.33E+00 & 2.58E+01\tabularnewline
30 & $10^{11}$ & 1.07E+01 & 2.39E+01 & 1.64E+01 & 4.72E+01 & 3.79E+01 & 8.24E+01\tabularnewline
\midrule
\multicolumn{8}{|c|}{\rule{0pt}{12pt} R$_{\text{EC}}$(QRPA/IPM) } \\[0.4ex]
\midrule
3 & $10^{3}$ & 5.61E+00 & 1.79E-02 & 1.34E+02 & 6.95E-01 & 5.77E+02 & 1.00E+00\tabularnewline
5 & $10^{3}$ & 5.78E+00 & 8.93E-01 & 1.07E+01 & 6.65E-01 & 1.02E+02 & 2.10E+00\tabularnewline
10 & $10^{3}$ & 2.88E+00 & 4.40E+00 & 2.05E+00 & 1.17E+00 & 1.60E+01 & 2.70E+00\tabularnewline
30 & $10^{3}$ & 3.58E+00 & 5.65E+00 & 3.78E+00 & 5.48E+00 & 1.08E+01 & 8.81E+00\tabularnewline
 &  &  &  &  &  &  & \tabularnewline
3 & $10^{7}$ & 5.53E+00 & 1.79E-02 & 1.33E+02 & 6.90E-01 & 5.75E+02 & 1.00E+00\tabularnewline
5 & $10^{7}$ & 5.78E+00 & 8.95E-01 & 1.07E+01 & 6.64E-01 & 1.02E+02 & 2.10E+00\tabularnewline
10 & $10^{7}$ & 2.88E+00 & 4.40E+00 & 2.04E+00 & 1.17E+00 & 1.60E+01 & 2.70E+00\tabularnewline
30 & $10^{7}$ & 3.59E+00 & 5.65E+00 & 3.77E+00 & 5.47E+00 & 1.08E+01 & 8.83E+00\tabularnewline
 &  &  &  &  &  &  & \tabularnewline
1 & $10^{11}$ & 3.76E-01 & 4.71E-01 & 3.94E-01 & 7.10E-01 & 1.82E-01 & 1.06E+00\tabularnewline
1.5 & $10^{11}$ & 3.82E-01 & 4.75E-01 & 3.91E-01 & 7.57E-01 & 1.81E-01 & 1.21E+00\tabularnewline
2 & $10^{11}$ & 3.89E-01 & 5.00E-01 & 3.90E-01 & 7.85E-01 & 1.87E-01 & 1.35E+00\tabularnewline
3 & $10^{11}$ & 4.05E-01 & 5.93E-01 & 3.94E-01 & 8.20E-01 & 2.22E-01 & 1.54E+00\tabularnewline
5 & $10^{11}$ & 4.32E-01 & 8.18E-01 & 4.11E-01 & 8.97E-01 & 3.59E-01 & 1.80E+00\tabularnewline
10 & $10^{11}$ & 5.38E-01 & 1.35E+00 & 5.86E-01 & 1.33E+00 & 1.18E+00 & 2.55E+00\tabularnewline
30 & $10^{11}$ & 2.78E+00 & 4.76E+00 & 2.94E+00 & 5.40E+00 & 7.43E+00 & 8.97E+00\tabularnewline
 \bottomrule
\end{tabular}}}
\end{table}

\begin{table}[pt]
\caption{\small Same as in Table~\ref{table:table4}, but for $\beta^{-}$-decay rates.}\label{table:table5} \hspace{-1cm}
\setlength{\aboverulesep}{0pt} \setlength{\belowrulesep}{0pt}
{\small
\centering {%\resizebox{!}{5cm}{%
\begin{tabular}{|cccccccc|}%{\textwidth}{c @{\extracolsep{\fill}} ccccccc}
\toprule
{\rule{0pt}{13pt} T$_{9}$} & $\rho$ Y$_{e}$ & $^{55}$Mn & $^{56}$Mn & $^{57}$Mn & $^{58}$Mn & $^{59}$Mn & $^{60}$Mn \\[0.4ex]
\midrule
\midrule
\multicolumn{8}{|c|}{\rule{0pt}{12pt} R$_{\beta^{-}}$(QRPA/LSSM) } \\[0.4ex]
\midrule
1 & $10^{3}$ & 1.99E+03 & 3.89E-02 & 8.02E-01 & 4.47E+00 & 2.70E+00 & 8.73E-01\tabularnewline
1.5 & $10^{3}$ & 3.18E+01 & 5.62E-02 & 9.68E-01 & 5.47E+00 & 2.90E+00 & 1.33E+00\tabularnewline
2 & $10^{3}$ & 2.83E+00 & 1.01E-01 & 1.11E+00 & 5.86E+00 & 3.05E+00 & 1.87E+00\tabularnewline
3 & $10^{3}$ & 1.93E-01 & 2.08E-01 & 1.44E+00 & 5.62E+00 & 3.32E+00 & 3.12E+00\tabularnewline
5 & $10^{3}$ & 5.83E-02 & 2.78E-01 & 1.25E+00 & 4.38E+00 & 3.02E+00 & 5.52E+00\tabularnewline
10 & $10^{3}$ & 8.24E-02 & 5.09E-01 & 3.54E-01 & 4.05E+00 & 1.48E+00 & 1.04E+01\tabularnewline
30 & $10^{3}$ & 1.59E-01 & 1.11E+00 & 5.02E-01 & 8.13E+00 & 1.98E+00 & 3.10E+01\tabularnewline
 &  &  &  &  &  &  & \tabularnewline
1 & $10^{7}$ & 1.14E+01 & 1.81E-03 & 8.30E-01 & 3.91E+00 & 2.74E+00 & 8.36E-01\tabularnewline
1.5 & $10^{7}$ & 1.44E+00 & 1.23E-02 & 1.01E+00 & 4.84E+00 & 2.94E+00 & 1.28E+00\tabularnewline
2 & $10^{7}$ & 3.76E-01 & 3.80E-02 & 1.16E+00 & 5.24E+00 & 3.08E+00 & 1.82E+00\tabularnewline
3 & $10^{7}$ & 1.17E-01 & 1.18E-01 & 1.49E+00 & 5.13E+00 & 3.36E+00 & 3.04E+00\tabularnewline
5 & $10^{7}$ & 5.55E-02 & 2.30E-01 & 1.23E+00 & 4.16E+00 & 3.02E+00 & 5.45E+00\tabularnewline
10 & $10^{7}$ & 8.09E-02 & 5.02E-01 & 3.48E-01 & 4.01E+00 & 1.47E+00 & 1.04E+01\tabularnewline
30 & $10^{7}$ & 1.59E-01 & 1.11E+00 & 5.01E-01 & 8.13E+00 & 1.98E+00 & 3.10E+01\tabularnewline
 &  &  &  &  &  &  & \tabularnewline
10 & $10^{11}$ & 1.46E-02 & 7.89E-02 & 4.55E-02 & 3.57E-01 & 1.57E-01 & 1.17E+00\tabularnewline
30 & $10^{11}$ & 5.78E-02 & 4.94E-01 & 2.00E-01 & 3.55E+00 & 8.32E-01 & 1.48E+01\tabularnewline
\midrule
\multicolumn{8}{|c|}{\rule{0pt}{12pt} R$_{\beta^{-}}$(QRPA/IPM) } \\[0.4ex]
\midrule
1 & $10^{3}$ & 8.09E+01 & 2.62E-02 & 7.01E-01 & 3.10E+00 & 2.95E+00 & 7.23E+00\tabularnewline
1.5 & $10^{3}$ & 2.53E+00 & 4.13E-02 & 8.07E-01 & 3.17E+00 & 3.20E+00 & 7.52E+00\tabularnewline
2 & $10^{3}$ & 1.66E-01 & 7.59E-02 & 8.83E-01 & 3.17E+00 & 3.35E+00 & 8.24E+00\tabularnewline
3 & $10^{3}$ & 6.15E-03 & 1.71E-01 & 6.59E-01 & 3.21E+00 & 2.76E+00 & 9.77E+00\tabularnewline
5 & $10^{3}$ & 3.05E-03 & 3.52E-01 & 1.03E-01 & 2.51E+00 & 3.06E-01 & 7.31E+00\tabularnewline
10 & $10^{3}$ & 2.13E-02 & 1.75E-01 & 2.60E-02 & 5.92E-01 & 2.80E-02 & 1.37E+00\tabularnewline
30 & $10^{3}$ & 1.69E-01 & 3.56E-02 & 5.97E-02 & 2.36E-01 & 2.08E-02 & 4.51E-01\tabularnewline
 &  &  &  &  &  &  & \tabularnewline
1 & $10^{7}$ & 5.00E-01 & 1.32E-03 & 7.45E-01 & 2.74E+00 & 3.06E+00 & 6.87E+00\tabularnewline
1.5 & $10^{7}$ & 9.53E-02 & 9.75E-03 & 8.61E-01 & 2.85E+00 & 3.31E+00 & 7.18E+00\tabularnewline
2 & $10^{7}$ & 1.56E-02 & 3.06E-02 & 9.40E-01 & 2.87E+00 & 3.47E+00 & 7.93E+00\tabularnewline
3 & $10^{7}$ & 3.32E-03 & 1.03E-01 & 6.47E-01 & 2.98E+00 & 2.80E+00 & 9.44E+00\tabularnewline
5 & $10^{7}$ & 3.02E-03 & 3.02E-01 & 9.77E-02 & 2.40E+00 & 3.01E-01 & 7.16E+00\tabularnewline
10 & $10^{7}$ & 2.14E-02 & 1.72E-01 & 2.57E-02 & 5.85E-01 & 2.78E-02 & 1.36E+00\tabularnewline
30 & $10^{7}$ & 1.69E-01 & 3.55E-02 & 5.97E-02 & 2.36E-01 & 2.09E-02 & 4.50E-01\tabularnewline
 &  &  &  &  &  &  & \tabularnewline
10 & $10^{11}$ & 4.06E-02 & 1.07E-02 & 1.86E-02 & 3.72E-02 & 8.05E-03 & 1.25E-02\tabularnewline
30 & $10^{11}$ & 1.83E-01 & 1.89E-02 & 5.36E-02 & 1.27E-01 & 1.51E-02 & 1.49E-01\tabularnewline
 \bottomrule
\end{tabular}}}
\end{table}

\end{document}